\documentclass[lettersize,journal]{IEEEtran}
\usepackage{amsmath,amsfonts}
\usepackage{algorithmic}
\usepackage{algorithm}
\usepackage{array}
\usepackage[caption=false,font=normalsize,labelfont=sf,textfont=sf]{subfig}
\usepackage{textcomp}
\usepackage{stfloats}
\usepackage{url}
\usepackage{verbatim}
\usepackage{graphicx}
\usepackage[square, numbers]{natbib}
\usepackage{hyperref}
\usepackage{multirow}
\usepackage{booktabs}
\usepackage{lipsum}
\usepackage{enumitem}
\usepackage{stfloats} 
\usepackage{makecell}
\usepackage{float}
\begin{document}
\title{FGAS: Fixed Decoder Network-Based Audio Steganography with Adversarial Perturbation Generation}

\author{Jialin Yan, Yu Cheng\thanks {Jialin Yan, Yu Cheng and Zhaoxia Yin (corresponding author, e-mail: zxyin@cee.ecnu.edu.cn) are with the School of Communication \& Electronic Engineering, East China Normal University, Shanghai 200241, China. Yu Cheng is also with the Shanghai Innovation Institute.}, Zhaoxia Yin,~\IEEEmembership{Member,~IEEE}, Xinpeng Zhang,~\IEEEmembership{Senior Member,~IEEE}\thanks {Xinpeng Zhang is with the School of Computer Science, Fudan University, Shanghai 200433, China.}, Shilin Wang,~\IEEEmembership{Senior Member,~IEEE}, Tanfeng Sun,~\IEEEmembership{Senior Member,~IEEE}, Xinghao Jiang,~\IEEEmembership{Senior Member,~IEEE}\thanks {Shilin Wang, Tanfeng Sun and Xinghao Jiang are with the School of
Electronic Information and Electrical Engineering, Shanghai Jiao Tong University, Shanghai 200240, China}}      

\maketitle
\begin{abstract}
The rapid development of Artificial Intelligence Generated Content (AIGC) has made high-fidelity generated audio widely available across the Internet, driving the advancement of audio steganography. Benefiting from advances in deep learning, current audio steganography schemes are mainly based on encoder-decoder network architectures. While these methods achieve promising perceptual quality, they typically require network pre-training, and maintaining anti-steganalysis performance at high payloads remains challenging. To address these issues, we propose Fixed Decoder Network-Based Audio Steganography with Adversarial Perturbation Generation (FGAS). Adversarial perturbations carrying a secret message are embedded into the cover audio to generate stego audio. The sender and receiver independently construct the same fixed decoder from a shared architecture and initialization key, avoiding the transfer of trained decoder parameters. In FGAS, we design a lightweight fixed decoder and propose an Audio Adversarial Perturbation Generation (A\textsuperscript{2}PG) strategy with an optional robust extension. With the decoder parameters frozen, A\textsuperscript{2}PG optimizes bounded perturbations for message recovery while keeping the stego audio perceptually and statistically close to the cover audio. Experimental results show that FGAS improves stego audio quality, achieving an average PSNR gain of over 10 dB compared to SOTA methods. The robustness-enhanced variant further improves message recovery under common audio processing attacks. Moreover, FGAS exhibits competitive anti-steganalysis performance across different relative payloads; under high-capacity embedding, it achieves a classification error rate approximately 2 percentage points higher than competing SOTA methods.
\end{abstract}
\section{Introduction}
\IEEEPARstart {S}{teganography} serves as a critical technique for information hiding \cite{11088141, DBLP:journals/sigpro/ChengCG25} and covert communication \cite{26,pan2025rethinking,guan2025non}, protecting individual privacy by embedding secret information within digital media \cite{DBLP:conf/icmcs/LiLZTH25,cheng2025robust,hu2024establishing,yang2023semantic,meng2023robust}.  The rapid proliferation of high-fidelity audio generated by AIGC has inherently provided an ideal and ubiquitous cover medium for covert communication. Consequently, audio steganography has resurfaced as a critical and challenging research area demanding in-depth investigation.\\
\begin{figure}[t]
\centering
\includegraphics[width=8.7cm]{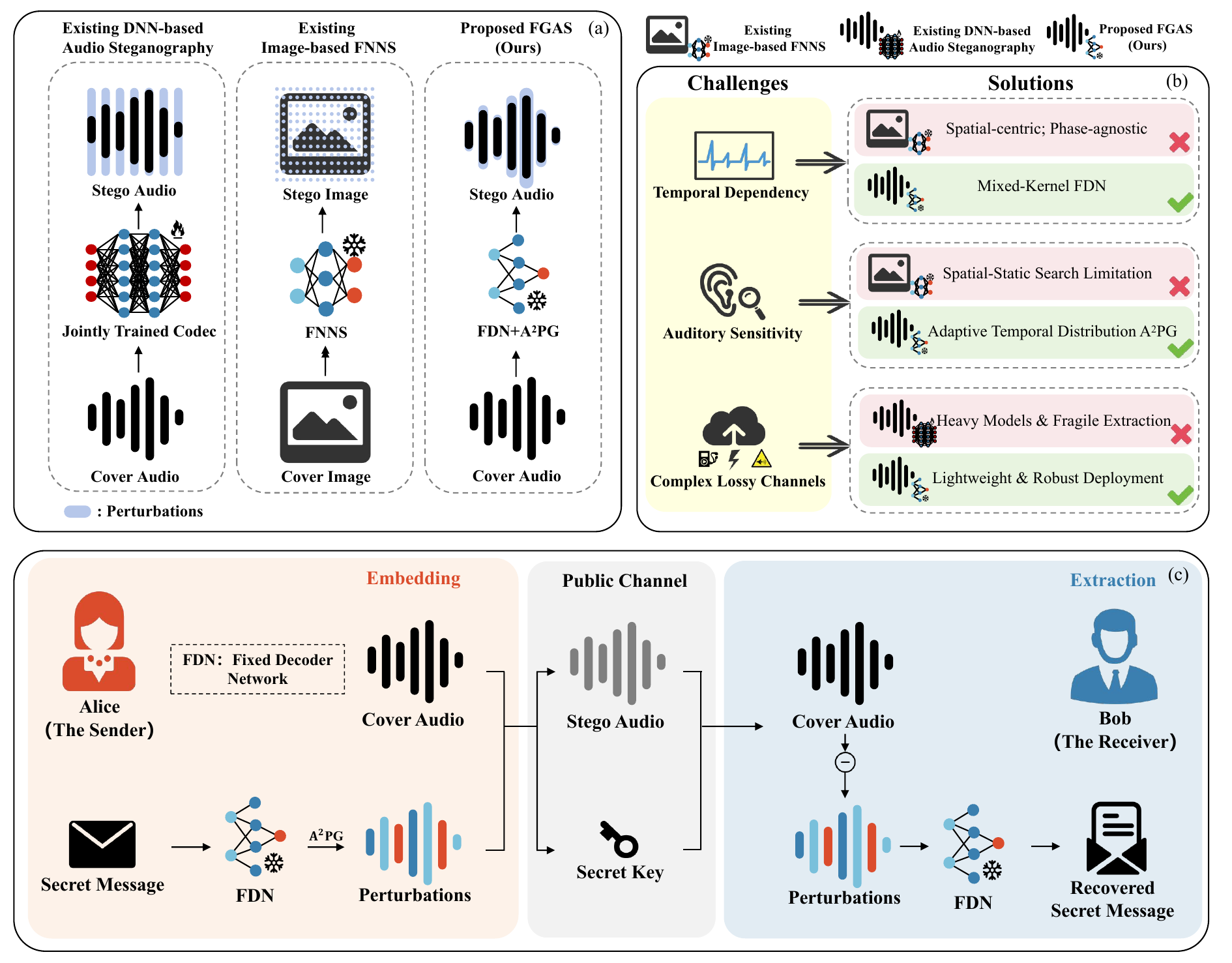}
\setlength{\abovecaptionskip}{0pt} 
\caption{Motivations and the proposed framework of FGAS: (a) The comparison between existing image-based FNNS, existing DNN-based audio steganography, and the proposed FGAS. (b) Key challenges in audio steganography and our corresponding solutions compared with existing methods. (c) The overall architecture of the proposed FGAS.}
\label{fig:1}
\end{figure}
\indent With the rapid maturation of multi-media steganalysis techniques \cite{DBLP:journals/jcst/WeiLTH25, DBLP:journals/tifs/WeiLH24, DBLP:journals/mms/WeiLLY23}, steganography faces increasingly rigorous security challenges \cite{DBLP:conf/ih/ZhouY0LW25,DBLP:journals/tifs/HuangLLTH24, DBLP:journals/tifs/GuanLZLZ23, DBLP:conf/icassp/YeHW024}. Traditional audio steganography, typically operating in the time domain~\cite{chen2019derivative,8,9,10}, relies on handcrafted embedding rules that often require extensive parameter tuning to maintain imperceptibility \cite{5,6,7}. While Deep Neural Network (DNN) based approaches~\cite{11,12,15,30} have enhanced anti-steganalysis performance~\cite{hu2025mutual,29}, they introduce significant deployment hurdles. Specifically, as illustrated in Fig.~\ref{fig:1} (b), existing methods suffer from the security risks and high communication overhead associated with transmitting large-scale pre-trained models, which can raise suspicion during transmission \cite{13,14}. To circumvent these issues, we propose Fixed Decoder Network-Based Audio Steganography (FGAS), as depicted in Fig.~\ref{fig:1} (c). In FGAS, the sender (Alice) and receiver (Bob) share a fixed decoder network that is initialized with a lightweight secret key. By sharing only the stego audio and the key, FGAS eliminates the need for massive model transmission, ensuring more secure and efficient covert communication.\\
\indent While this paradigm mitigates the risks associated with model transmission, designing a viable Fixed Decoder Network for audio is uniquely challenging due to the inherent characteristics of 1D waveforms. Unlike images \cite{17,18,22} with spatial redundancy, high-sampling-rate audio signals exhibit an expansive dynamic range, in which the Human Auditory System (HAS) is extremely sensitive to temporal discontinuities and spectral anomalies. Standard fixed-architecture networks \cite{16,19} often fail in this domain due to their inability to model long-range temporal dependencies, leading to synchronization issues and audible artifacts \cite{guo2025audio}. To address this, we design a temporal-feature-aware fixed decoder that captures broader temporal contexts and stabilizes feature distributions, ensuring reliable message extraction. While the decoder remains fixed, we propose an Audio Adversarial Perturbation Generation (A\textsuperscript{2}PG) strategy to iteratively optimize size-constrained perturbations, guiding the stego audio to trigger the desired output while remaining indistinguishable to steganalyzers. Furthermore, we introduce a Robustness-Enhanced A\textsuperscript{2}PG (A\textsuperscript{2}PG-R). By employing an adversarial curriculum learning approach, A\textsuperscript{2}PG-R enables the perturbations to withstand common channel distortions, thereby guaranteeing generalized robustness in lossy environments.\\
\indent  We conduct extensive experiments across speech and music datasets to comprehensively validate FGAS's performance. Our standard FGAS configuration demonstrates superior imperceptibility, achieving an average Peak Signal-to-Noise Ratio (PSNR) more than 10 dB higher than competing methods, complemented by high scores in perceptual metrics such as Perceptual Evaluation of Audio Quality (PEAQ) \cite{thiede2000peaq}. In terms of anti-steganalysis performance, FGAS successfully evades detection by sophisticated deep learning steganalyzers  \cite{23,24} and traditional feature-based detectors  \cite{luo2018improved}, maintaining near-random-guess accuracy across various payloads. Furthermore, we introduce the A\textsuperscript{2}PG-R as an optional extension. The results demonstrate that the Curriculum Learning-based A\textsuperscript{2}PG-R mechanism effectively fortifies the resilience of stego audio against lossy channels. By prioritizing channel survivability, this approach achieves a superior balance between embedding fidelity and robust data extraction. These findings affirm the practical value of FGAS as a secure, lightweight, and adaptable framework for robust covert audio communication.\\
\indent Our main contributions are summarized below:
\begin{itemize}[label={•},labelindent=2em,leftmargin=*,itemindent=0em]
    \item The Fixed Decoder Network-Based Audio Steganography (FGAS) is proposed, which is the first scheme to employ a fixed decoder in the audio domain. Its Fixed Decoder Network reliably extracts bits from 1D audio and eliminates the need to transmit large model parameters, enabling lightweight, secure covert communication.
    \item An Audio Adversarial Perturbation Generation (A\textsuperscript{2}PG) with an optional A\textsuperscript{2}PG-R extension is designed. A\textsuperscript{2}PG guides perturbation directions to keep stego audio close to the cover under steganalyzer constraints, while A\textsuperscript{2}PG-R applies curriculum learning to enhance robustness to channel distortions.
    \item Extensive experiments on speech and music demonstrate the practical advantages of FGAS. Our method delivers higher stego-audio quality, more accurate secret-message recovery, strong robustness against common channel distortions, and superior anti-steganalysis performance against both neural and traditional detectors.
\end{itemize}

 The remainder of this paper is organized as follows. Section II provides a brief overview of traditional and DNN-based time-domain audio steganography and of image steganography using a fixed neural network approach. Section III describes the proposed steganography scheme, including the A\textsuperscript{2}PG strategy and the perturbation-sensitive fixed decoder network design. The experimental results and analysis are given in Section IV. Finally, Section V gives the conclusion.\\
\section{Related Work}
\indent In this section, we divide existing audio steganography into two categories, traditional audio steganography and DNN-based audio steganography. In addition, the application and development of fixed neural networks is investigated in steganography.
\subsection{Traditional Audio Steganography}
\indent Traditional audio steganography techniques rely on meticulously hand-crafted embedding schemes that introduce imperceptible modifications into the cover audio. Early efforts focused on improving Least Significant Bit (LSB) hiding by enhancing capacity and reducing distortion, exemplified by methods like the high bit rate LSB watermarking of Roy et al. \cite{5}, the Dual Randomness LSB scheme by Vimal et al. \cite{6}, and the Binaries of Message Size Encoding (BMSE) approach proposed by Mahmoud et al. \cite{7}. However, these basic LSB schemes inherently suffer from unsatisfactory detection performance against modern steganalyzers, prompting a shift toward adaptive distortion modeling. More sophisticated methods were introduced to enhance security by defining and controlling embedding costs: Chen et al. \cite{chen2019derivative} utilized a Derivative-based Steganographic Distortion (DFR) that leverages audio residuals; Luo et al. \cite{8} proposed an adaptive scheme combining Advanced Audio Coding (AAC) and Syndrome Trellis Codes (STCs) \cite{9} based on compression residuals; and Su et al. \cite{10} introduced an approach leveraging Micro-Amplitude Suppression (MAS) and Generalized Audio Intrinsic Energy (GAIE) to stabilize adaptive costs. Despite these advances in hand-crafted security, designing schemes that achieve high capacity while resisting advanced steganalysis remains a critical challenge.\\
\indent Although requiring extensive manual calibration to preserve imperceptibility, these techniques typically exhibit diminishing stealth capabilities as payload sizes increase, thus becoming more susceptible to contemporary steganalyzers \cite{28}.\\
\subsection{DNN-Based Audio Steganography}
\indent Due to the rapid development of deep learning, complementary to hand-crafted schemes, several deep learning-based steganography methods have emerged. These can be broadly divided into embedding-based and generative approaches. Embedding methods leverage neural networks to optimize payload placement: for instance, Yang et al. \cite{11} used a GAN to learn adaptive probabilities, Wu et al. \cite{12} employed Iterative Adversarial Attacks (IAA) for anti-steganalysis improvement, Gelet et al. \cite{13} used STDCT spectrograms to hide images, and Zhang et al. \cite{11007018} proposed HIFI-Stego by embedding messages into a decoupled content vector for high-fidelity security. Alternatively, generative approaches aim to bypass direct embedding: Chen et al. \cite{30} proposed carrier reproducible steganography, while distribution-preserving systems \cite{14} and coverless methods \cite{15} rely on generative models and mapping rules. Despite the substantial gains these schemes offer in security and imperceptibility, a crucial, shared limitation is the heavy reliance on extensive datasets and significant computational resources for effective training.
\subsection{Fixed Neural Network-based Steganography}
To avoid network training and transmission, Fixed Neural Network Steganography (FNNS) emerged, utilizing a fixed decoder network for data hiding and recovery. Established in image processing, methods include Kishore et al. \cite{16} (capacity enhancement) and Cover-separable FNNS (Cs-FNNS) \cite{18}. More recently, Robust FNNS (RFNNS) \cite{22} improved security using texture-aware localization.\\
\indent However, these image-centric designs are ill-suited for audio. The high-sampling-rate 1D nature of audio, combined with the Human Auditory System's extreme sensitivity to spectral anomalies, demands far more stringent SNR constraints than for 2D images. Moreover, existing fixed-architecture models struggle to capture long-range temporal dependencies, often leading to desynchronization in extraction and audible artifacts \cite{guo2025audio}. This gap necessitates a specialized FNNS approach tailored to the unique physiological and structural properties of audio signals.\\
\section{The Proposed Method}
\begin{figure*}[t]
\centering
\includegraphics[width=18cm]{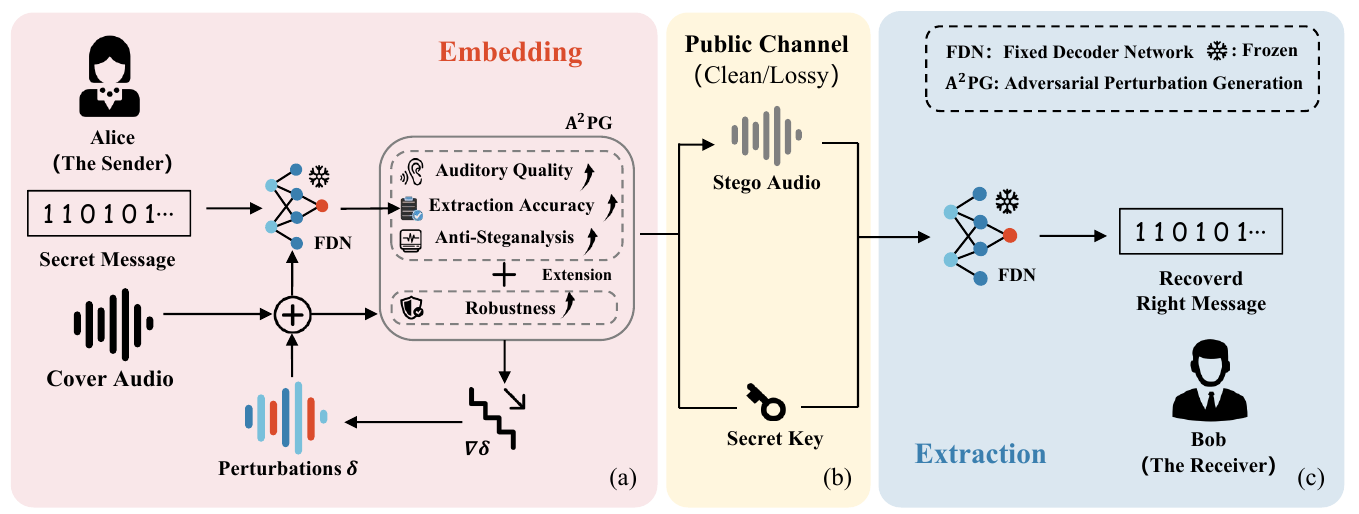}
\setlength{\abovecaptionskip}{0pt} 
\caption{FGAS scheme: (a): Alice (The Sender) uses the proposed audio Adversarial Perturbation Generation (A\textsuperscript{2}PG) strategy to generate a perturbation, which can be decoded to the secret message by a fixed neural network decoder, then adds the perturbation on cover audio (controlled by a key) to produce a stego audio. (b): Stego audio is transmitted over the public channel. (c): Bob (The receiver) decodes the stego audio using a shared key and the same decoder network to obtain the secret message.}
\label{fig:2}
\end{figure*}
\indent This section begins with an overview of the proposed framework, followed by a detailed explanation of the audio Adversarial Perturbation Generation (A\textsuperscript{2}PG) strategy. Finally, we describe the decoder network design.

\subsection{Framework of the Proposed Scheme}

\indent The proposed audio steganography scheme is built upon a fixed neural network architecture. Let $A_c$ denote the lossless cover audio in WAV format, characterized by $T$ channels and $H$ samples, with amplitudes normalized to $[-1,1]$. The secret message to be transmitted, denoted as $M$, is a binary bit stream of length $K$. For clarity, the relevant symbols used throughout this paper are summarized in Table~\ref{tab:1}.

\begin{table}[!t]
\centering
\caption{Notations}
\label{tab:1}
\begin{tabular}{cc}
\toprule
Notations & Description \\
\midrule
$A_c$ & Cover audio $A_c \in [-1,1]^{T \times H}$ \\
$A_s$ & Stego audio $A_s \in [-1,1]^{T \times H}$ \\
$\delta$ & Perturbation $\delta \in [-\varepsilon,\varepsilon]^{T \times H}$ \\
$M$ & Secret message $M \in \{0,1\}^K$ \\
$D_e(\cdot)$ & Fixed decoder $D_e: \mathbb{R}^{T \times H} \to [0,1]^K$ \\
\bottomrule
\end{tabular}
\end{table}

\indent According to the framework depicted in Fig.~\ref{fig:2}, the pipeline comprises two primary phases: message embedding at the sender's side and message extraction at the receiver's side. On the sender side, for a given cover audio $A_c$ and secret message $M$, Alice generates a message-carrying perturbation $\delta$ using the proposed Audio Adversarial Perturbation Generation (A\textsuperscript{2}PG) strategy. A\textsuperscript{2}PG iteratively optimizes the perturbation to balance audio fidelity, message recovery, and anti-steganalysis performance. The resulting stego audio is constructed as $A_s=A_c+\delta$, subject to the perturbation bound and the valid audio amplitude range. The decoder weights are initialized using a shared key $k_r$ and remain fixed throughout perturbation optimization.

\indent On the receiver side, Bob independently reconstructs the same fixed decoder $D_e$ using the shared network architecture and initialization key $k_r$. The perturbation $\delta$ is obtained by subtracting the original cover audio $A_c$ from the received stego audio $A_s$, and is subsequently fed into the fixed decoder to recover the secret message $M$. The extraction process is expressed as follows:
\begin{equation}
\label{eq:1}
D_e(\delta)=M.
\end{equation}
\subsection{Audio Adversarial Perturbation Generation strategy}
\indent In the context of propagating on public channels, it is imperative to consider the potential for attackers with ulterior motives to detect the intercepted stego audio using advanced steganalyzers and attempt to extract the secret information. In this scenario, the concealment of the stego audio and its anti-steganalysis performance are particularly important. To address the aforementioned issues, an audio adversarial perturbation generation (A\textsuperscript{2}PG) strategy is proposed. A\textsuperscript{2}PG strategy aims to find a perturbation $\delta \in [-\epsilon, \epsilon]^{T \times H}$ that satisfies the following three properties: 1) Minimizing the difference between cover audio $A_c$ and stego audio $A_s$ to ensure auditory covert; 2) ensuring that fixed decoder network $D_e(\cdot)$ can decode secret messages $M$; 3) misleading the target stegalyzer to misclassify the stego audio as cover audio to enhance anti-steganalysis performance. The optimization problem mentioned above can be expressed by the following mathematical formulation:\\
\begin{equation}
\label{eq:fgas_opt}
\min_{\delta}\  dist(A_c, A_s)
\end{equation}
\begin{equation}
\label{eq:2}
\left.\mathrm{s.t.}\quad\left\{
\begin{array}
{l}\|A_c-A_s\|\leq\epsilon\\
D_e(\delta)=M \\
J(A_s) \leq \tau\\
-1\leq{A_s}\leq1
\end{array}\right.\right.,
\end{equation}
where the first condition strictly limits the size of the perturbation $\delta$ to ensure its imperceptibility. 
$J(\cdot)$ denotes a set of deep learning-based steganalyzers \cite{23, 24}, each of which takes an arbitrary audio signal as input and outputs the probability of the presence of a hidden message. 
The third constraint $J(A_s) \leq \tau$ ensures that the detection probability remains below a sufficiently small threshold $\tau$, indicating that the stego audio successfully deceives the steganalyzers. The final constraint ensures that the amplitude of the generated stego audio $A_s$ remains within the normalized range of $[-1, 1]$.\\
\indent To ensure that the generated perturbation $\delta$ simultaneously satisfies the aforementioned constraints, we formulate the embedding process as a multi-objective optimization problem. By iteratively minimizing a joint loss function, the perturbation $\delta$ is refined to achieve an optimal balance between stego audio quality, message recovery accuracy, and anti-steganalysis performance.\\
\indent Correspondingly, to satisfy the first constraint above and mitigate the impact of perturbation on the quality of the cover audio, the perturbation introduced during the embedding process should be as minimal as possible. A loss function is employed as follows:
\begin{equation}
\label{eq:3}
L_1=MSELoss(As,Ac).
\end{equation}
$L_1$ measures the difference between the stego audio and the cover audio. Specifically, to constrain the perturbation within the limits, $\delta$ is bounded by $\epsilon$, as shown in the following Eq.~\ref{eq:4}.
\begin{equation}
\label{eq:4}
-\epsilon\leq\delta\leq \epsilon\\
\end{equation}
\indent In addition to maintaining the quality of stego audio, it is critical to reliably extract the secret message. In order to ensure the accuracy of the recovery information and to satisfy the second constraint, the binary cross-entropy (BCE) loss function is introduced:
\begin{equation}
\label{eq:5}
L_2=BCELoss(M,D_e(\delta)).
\end{equation}
$L_2$ measures the difference by calculating the cross entropy between the extracted bits of secret message and the original secret message, to maximize the precision of the recovery of secret message.\\
\indent To actively fool the steganalyzers, satisfy the third condition in the optimization task, a paired input batch is constructed:
\begin{equation}
\label{eq:6}
Z = \begin{bmatrix} A_c, A_s \end{bmatrix}, \quad Y = [0, 1]^\top,
\end{equation}
where label “0” denotes the cover signal $A_c$ and “1” denotes the stego signal $A_s$. Pre-trained steganalyzers \cite{23,24} and traditional feature-based detectors \cite{luo2018improved} are incorporated into the later iterations to provide gradient feedback for perturbation refinement, thereby enhancing the anti-steganalysis performance of the generated stego audio.\\
\indent Let the raw outputs from steganalyzers for the input batches $Z = \begin{bmatrix} A_c, A_s \end{bmatrix}$ be $s(Z)=[s_0,s_1]^\top$, and the corresponding predicted probabilities $p_j$ (Softmax) be:
\begin{equation}
\label{eq:7}
p_j=\frac{\exp(s_j)}{\exp(s_0)+\exp(s_1)},\quad j\in\{0,1\}.
\end{equation}
\indent The adversarial detection loss is then defined as:
\begin{equation}
\label{eq:8}
L_3=CrossEntropyLoss(Z,Y),
\end{equation}
\begin{equation}
\label{eq:9}
CrossEntropyLoss= -\sum_{j=0}^1 Y_j \log \left( \frac{e^{s_j}}{e^{s_0} + e^{s_1}} \right).
\end{equation}
$L_3$ measures the discrepancy between the predicted probability distribution and the ground-truth label distribution, aiming to mislead the steganalyzer into misclassifying stego audio as cover.\\
\indent On this basis, we integrate the three objectives: stego audio quality, message recovery accuracy, and anti-steganalysis performance into a unified optimization framework. The final joint loss function $L$ is formulated as:
\begin{equation}
\label{eq:10}
L=\alpha{L_1}+\beta{L_2}+\gamma{L_3},
\end{equation}
where $\alpha$, $\beta$, and $\gamma$ are hyperparameters that balance the contributions of different loss functions.\\
\begin{figure*}[t]
\centering
\includegraphics[width=18cm]{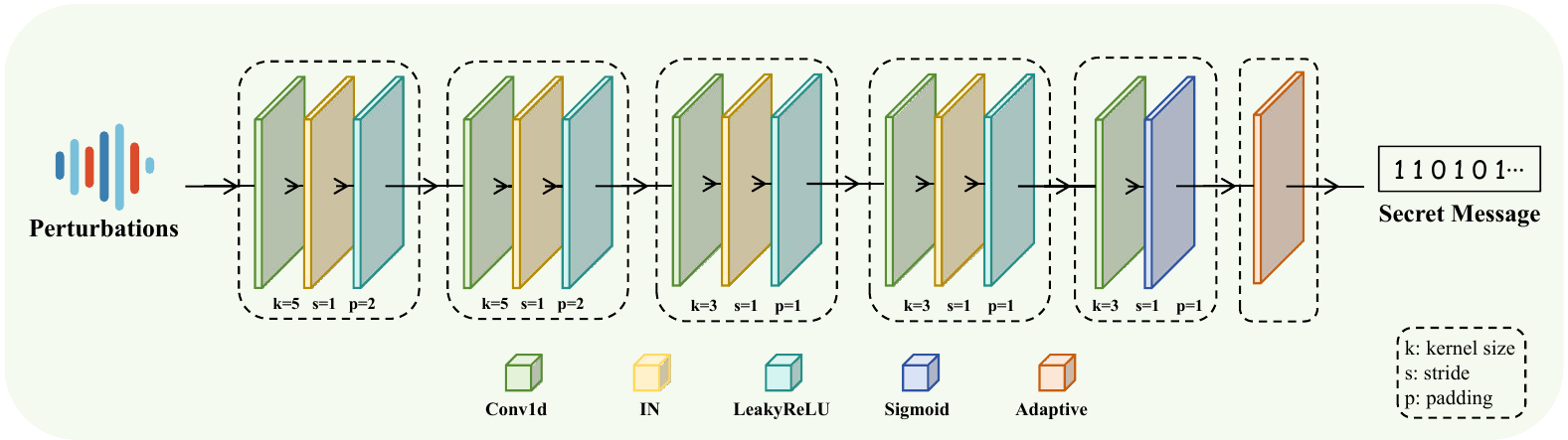}
\setlength{\abovecaptionskip}{0pt} 
\caption{Detailed structure of fixed decoder network and the process of perturbation decoding into secret message.}
\label{fig:3}
\end{figure*}
\subsection{Robustness-Enhanced A\textsuperscript{2}PG}
While the fundamental \(\mathrm{A}^2\mathrm{PG}\) strategy provides high
stego-audio quality, strong anti-steganalysis performance, and accurate
message extraction in clean channels, real-world transmission may
introduce distortions such as lossy compression, environmental noise,
and resampling. These distortions can disrupt the perturbation and reduce
message recovery accuracy. To improve robustness, robustness-enhanced
\(\mathrm{A}^2\mathrm{PG}\) (\(\mathrm{A}^2\mathrm{PG}\)-R) incorporates
channel attacks into perturbation optimization using a curriculum
schedule. Specifically, the clean-channel message recovery loss \(L_2\)
in Eq.~\ref{eq:5} is replaced with a robustness loss \(L_r\) that accounts
for message recovery under both clean and attacked conditions.
The resulting objective is
\begin{equation}
\label{eq:apg-r-total}
L = \alpha L_1 + \beta L_r + \gamma L_3,
\end{equation}
where \(L_1\) and \(L_3\) denote the audio distortion loss and
anti-steganalysis loss, respectively. Let \(\mathcal{A}(\cdot)\) denote
the channel attack applied to the stego audio \(A_s\). The attacked audio
is \(A_{\mathrm{att}}=\mathcal{A}(A_s)\), and its residual relative to
the original cover audio is \(\delta_{\mathrm{att}}=A_{\mathrm{att}}-A_c\).
The robustness loss is
\begin{equation}
\label{eq:robustness-term}
\begin{split}
L_r ={}& \omega\,\operatorname{BCE}\bigl(M,D_e(\delta)\bigr) \\
       &+ (1-\omega)\,\operatorname{BCE}
          \bigl(M,D_e(\delta_{\mathrm{att}})\bigr),
\end{split}
\end{equation}
where \(M\) is the target message and \(\operatorname{BCE}\) denotes
binary cross-entropy. For clean-channel optimization, \(\omega=1\),
so only the clean recovery term contributes to \(L_r\).

The \(\mathrm{A}^2\mathrm{PG}\)-R curriculum comprises 13 stages. It begins with progressively stronger Gaussian noise,
then increases MP3 compression strength by lowering the bitrate.
Subsequent stages introduce AAC compression, low-pass and band-pass
filtering, and finally time stretching. The perturbation is carried
forward between stages. At iteration \(i\), the stage index is
\(s=\lceil i/1000\rceil\), and the clean-channel weight is held fixed
within that stage at \(\omega_s=0.8-0.05(s-1)\). Thus, the clean-channel
weight decreases linearly from \(0.8\) to \(0.2\) over the curriculum,
while the weight on attacked-channel recovery increases from \(0.2\)
to \(0.8\). The attack operators are described below.

\begin{itemize}
    \item \textbf{Gaussian noise.}
    Gaussian noise is added to \(A_s\) at a prescribed signal-to-noise
    ratio (SNR). For a cover signal with mean power \(P_c\), the noise
    standard deviation is
    \(\sigma=\sqrt{P_c\,10^{-\mathrm{SNR}/10}}\).

    \item \textbf{MP3 and AAC compression.}
    Both attacks approximate spectral quantization and high-frequency
    suppression. For MP3, the short-time Fourier transform magnitude
    of \(A_s\) is quantized to discrete levels. For AAC, the simulation
    uses the modified discrete cosine transform followed by nonuniform
    quantization based on a psychoacoustic model. In both cases,
    frequencies above a bitrate-dependent cutoff are suppressed to model
    the bandwidth limitations of the codec.

    \item \textbf{Filtering.}
    Differentiable low-pass and band-pass filters are applied to \(A_s\)
    to simulate frequency-selective attenuation.

    \item \textbf{Time stretching.}
    The stego audio \(A_s\) is resampled from its original length \(H\)
    to a target length \(H'=kH\), where \(k\) is the stretch factor,
    and then linearly interpolated back to \(H\).
\end{itemize}

Algorithm~\ref{alg:1} summarizes the optimization procedure for
\(\mathrm{A}^2\mathrm{PG}\) and \(\mathrm{A}^2\mathrm{PG}\)-R.
In the robustness-enhanced variant,
\(\operatorname{Schedule}(\mathcal{A},i)\) selects the attack for the
current stage, and the corresponding \(\omega_s\) is used in
Eq.~\ref{eq:robustness-term}.
\begin{algorithm}[t]
\caption{Audio Adversarial Perturbation Generation (A\textsuperscript{2}PG) with Robustness Extension}
\label{alg:1}
\begin{algorithmic}[1]
\REQUIRE Cover audio $A_c$, secret bits $M$, decoder $D_e(\cdot)$, detectors $\{J_k\}$,\\
\hspace{\algorithmicindent}attack set $\mathcal{A}$, hyperparameters $\alpha,\beta,\gamma,\epsilon,\eta, \omega$, iterations $N$
\ENSURE Perturbation $\delta$
\STATE $\delta \leftarrow \mathbf{0}$
\FOR{$i = 1$ \TO $N$}
  \STATE $A_s \leftarrow \operatorname{project}(A_c + \delta,\,-1,1)$ \hfill \COMMENT{Build stego audio}
  \STATE $M \leftarrow D_e(\delta)$\hfill \COMMENT{Standard decoding}
  \STATE \COMMENT{Robustness Component}
  \STATE $A_{att} \leftarrow \operatorname{Schedule}(\mathcal{A}, i)(A_s)$     \hfill \COMMENT {Curriculum attack}
  \STATE $ M \leftarrow D_e(\delta_{att})$\hfill \COMMENT{Robust decoding}
  \STATE \COMMENT{Optimization}
  \STATE $\ell_1 \leftarrow \|A_s -A_c\|_2^2$\hfill \COMMENT{Distortion loss}
  \STATE $\ell_{2} \leftarrow \omega \operatorname{BCE}(M, D_e(\delta)) + (1-\omega) \operatorname{BCE}(M, D_e(\delta_{att}))$,
   $A^2PG (\omega=1) or A^2PG-R (\omega < 1)$    \hfill \COMMENT{Recovery loss}
  \STATE $\ell_3 \leftarrow -\sum_{k=1}^K \log\bigl(1 - J_k(A_s)\bigr)$ \hfill \COMMENT{Anti-detection loss}
  \STATE $\ell \leftarrow \alpha\,\ell_1 + \beta\,\ell_{2} + \gamma\,\ell_3$ \hfill \COMMENT{Total loss}
  \STATE $\delta \leftarrow \operatorname{AdamStep}(\delta,\nabla_{\delta}\ell,\eta)$
  \STATE $\delta \leftarrow \operatorname{project}(\delta,\,-\epsilon,\epsilon)$ \hfill \COMMENT{Constraint}
\ENDFOR
\RETURN $\delta$
\end{algorithmic}
\end{algorithm}
\subsection{Fixed Decoder Network}
\indent The architecture of the decoder network $D_e(\cdot)$ serves as the pivotal framework for ensuring both accurate extraction of secret message and superior imperceptibility. Given that raw audio signals are high-sampling-rate 1D waveforms with an expansive dynamic range, they are highly sensitive to temporal discontinuities and spectral anomalies. To mitigate the risk of high-frequency artifacts perceptible to the HAS and to effectively capture long-term temporal dependencies within the signal, we develop a specialized, lightweight Fixed Decoder Network (FDN) specifically optimized for the inherent characteristics of 1D audio data.\\
\indent The decoder comprises a sequence of five Conv1D blocks, each employing convolution, Instance Normalization, and LeakyReLU activation. The design adopts a mixed kernel strategy, using larger kernels in the initial blocks to capture broad temporal contexts and smaller kernels in subsequent blocks to capture fine local details within the high-sampling-rate waveform. InstanceNorm1D normalizes each feature channel along the time axis within each audio sample to stabilize its scale. Each convolution uses unit stride and symmetric padding chosen to preserve the input length, so that fine-grained temporal features remain at the original resolution until pooling. Finally, adaptive average pooling maps the temporal response to a sequence of the target message length, and the final Sigmoid activation converts each pooled value into a bit probability. The detailed structure and decoding process are illustrated in Fig.~\ref{fig:3}.\\
\indent After determining the architecture of the decoder network $D_e(\cdot)$ and agreeing on the shared key $k_r$, both the sender and receiver can independently construct an identical decoder without transmitting trained model parameters. The decoder remains fixed and is reused across communications, while adaptation to each cover signal is handled through perturbation optimization at the sender, without decoder retraining or updates. Even with a one-time secure setup, this lightweight design keeps initial deployment overhead and ongoing receiver-side storage requirements low, supporting decoder reuse at resource-constrained receivers.
\section{Experiment}
\subsection{Experiment Setups}
\footnotetext[1]{https://catalog.ldc.upenn.edu/}
\footnotetext[2]{https://keithito.com/LJ-Speech-Dataset/}
\subsubsection{\bf{Implementation Details}}
We conduct experiments on four datasets: TIMIT\textsuperscript{1}, LJSpeech\textsuperscript{2}, GTZAN\textsuperscript{3}, and AudioSet\textsuperscript{4}. From each dataset, we collect 3,000 mono, 16-bit audio clips, resample them to 16~kHz, normalize their amplitudes to $[-1, 1]$, and standardize their duration to 1~s (16,000 samples). We store all clips in WAV format. We minimize the loss function using Adam~\cite{25} with an initial learning rate of 0.001. For the base method, we perform 2,000 optimization updates. For $\mathrm{A}^2\mathrm{PG}$-R, the curriculum comprises 13 stages with 1,000 optimization updates per stage. A pre-trained steganalytic network~\cite{23,24} provides gradients for perturbation refinement. We initialize the decoder weights using a random seed, keep them fixed, and secure them as the extraction key $k_r$.\\
\subsubsection{\bf{Evaluation}} The performance of the proposed method is evaluated across four dimensions, as detailed in Table~\ref{tab:evaluation}. Specifically, we employ PSNR and PEAQ to assess the objective fidelity and perceptual quality of stego audios, respectively. For security, the Classification Error Rate ($\overline{P}_{E}$) is used to quantify anti-steganalysis performance, with higher $\overline{P}_{E}$ indicating stronger resistance to detection. Message recovery and robustness are measured by the extraction accuracy of secret bits under clean and various attack conditions. For a fair comparison, all experiments are conducted under a constant embedding payload measured in bits per sample (bps).\\

\begin{table}[!t]
\centering
\caption{Performance Dimensions and Evaluation Metrics}
\label{tab:evaluation}
\begin{tabular}{ccc}
\toprule
Section & Evaluation Metrics& Index \\
\midrule
B.Robustness & Accuracy(\%)&Table~\ref{tab:robustnesswithout}\\
\makecell{C.Anti-Steganalysis\\Performance} & $\overline{P}_{E}$(\%)&\makecell{Table~\ref{tab:ChenNetResults}, Fig~\ref{tab:LinNetResults},\\Table~\ref{tab:TMFresults}, Table~\ref{tab:absteg}}\\
D.Stego Audio Quality & PSNR(dB) and PEAQ &Table~\ref{tab:2}, Table~\ref{tab:abquality} \\
\bottomrule
\end{tabular}
\end{table}
\begin{figure}[!t]
\centering
\includegraphics[width=9cm]{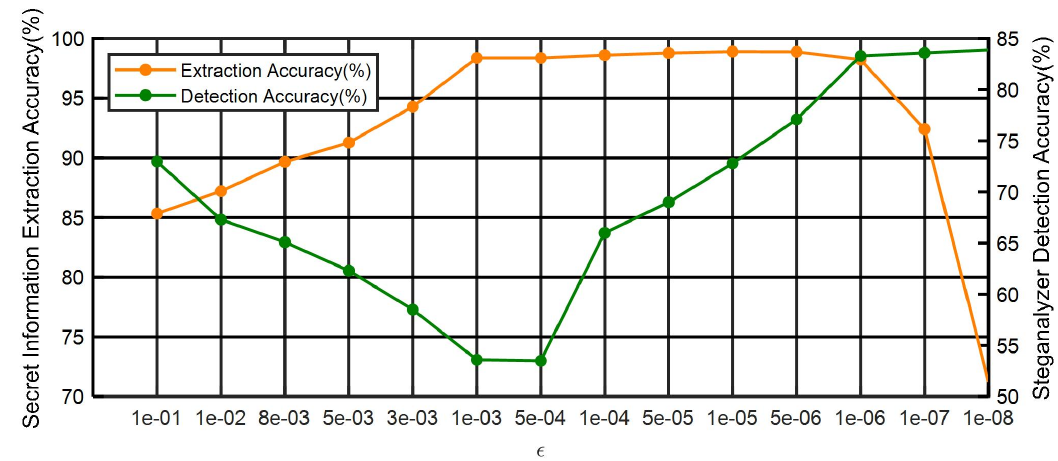}
\setlength{\abovecaptionskip}{0pt} 
\caption{The impact of $\epsilon$ on detection accuracy and recovery accuracy. Note that for $\epsilon \le 8\times 10^{-3}$, the PSNR exceeds 100 dB and PEAQ reaches 4.54, ensuring superior imperceptibility.}
\label{fig:eps}
\end{figure}
\subsubsection{\bf{Hyperparameters}} In the experiment, the hide loss weight $\alpha$ is set to 1, the recovery loss weight $\beta$ is set to 0.5, and the anti-steganalysis loss weight $\gamma$ is set to 0 for the first 1,500 iterations and dynamically increases to 0.01 for the last 500 iterations. \\
\indent We investigate how varying $\epsilon$ affects PSNR, detection accuracy, and message recovery performance, with other hyperparameters fixed. Experiments show that the audio quality index PSNR increases as $\epsilon$ decreases, and when $\epsilon$ is less than 8e-3, the PSNR exceeds 100dB, ensuring good listening quality. The variation of the steganalyzer detection accuracy and secret message recovery accuracy for different $\epsilon$ values is shown in Fig.~\ref{fig:eps}, with $\epsilon$ finally set to $0.001$.\\
\begin{table*}[t]
\caption{Secret message recovery accuracy (\%) of the tested steganography methods under six common attacks at a relative payload of 0.1 bps.}
\label{tab:robustnesswithout}
\centering
\small
\setlength{\tabcolsep}{3pt}
\begin{tabular}{@{}l*{15}{c}@{}}
\toprule
\multirow{2}{*}{Method}
& \multirow{2}{*}{Clean}
& \multicolumn{3}{c}{Noise}
& \multicolumn{5}{c}{MP3 compression (kbps)}
& AAC & LP-F & BP-F
& \multicolumn{3}{c}{Stretch} \\
\cmidrule(lr){3-5}   \cmidrule(lr){6-10}
\cmidrule(lr){11-11} \cmidrule(lr){12-12}
\cmidrule(lr){13-13} \cmidrule(lr){14-16}
& & 20 dB & 30 dB & 50 dB
& 16 & 32 & 64 & 128 & 256
& 256 kbps & 4k & 0.3--8k & 0.8 & 0.9 & 1.1 \\
\midrule
DFR \cite{chen2019derivative} & -
& 50.16 & 50.21 & 49.87
&50.06 &50.17 & 50.17 & 50.17 & 50.17
& 49.22 & 50.03 & 49.61 & 50.12 & 49.92 & 50.04 \\
\midrule
ACC \cite{8} & -
& 50.03 & 50.11 & 50.02
&48.99 & 50.26& 49.17 & 50.11 & 48.86
& 50.01 & 50.02 & 49.33 & 50.06 & 49.82 & 49.04 \\
\midrule
IAA\_flat \cite{12} & -
& 49.38 & 50.01 & 51.37
&49.43 & 48.75& 50.16 & 50.03 & 50.07
& 50.13 & 50.03 & 50.01 & 50.22 & 50.14 & 49.04 \\
\midrule
GAIE-MAS \cite{10} & -
& 48.73 & 50.12 & 51.33
&50.02 &50.01 & 50.02 & 50.02 & 50.02
& 49.90 & 50.27 & 48.66 & 50.13 & 50.44 & 49.98 \\
\midrule
Hifi-Stego \cite{11007018} & 99.62
& 60.14 & 72.53 & 88.67
&49.40 &48.23 & 52.03 & 64.80 & 72.37
& 53.61 & 51.09 & 50.36 & 51.22 & 51.25 & 50.33 \\
\midrule
FGAS (Ours) & 99.99
& \textbf{79.12} & \textbf{91.31} & \textbf{94.44}
& \textbf{66.26} & \textbf{78.03} & \textbf{88.75} & \textbf{89.69} & \textbf{89.75}
& \textbf{87.44} & \textbf{65.25} & \textbf{74.12}
& \textbf{94.18} & \textbf{94.99} & \textbf{93.44} \\
\bottomrule
\end{tabular}
\end{table*}
\begin{table*}[t]
\centering
\caption{The Anti-Steganalysis Performance $\overline{P}_{E}$ (IN \%) of The Tested Steganography methods in Resisting The Detection of \textbf{ChenNet} Steganalyzer}
\label{tab:ChenNetResults}
\begin{tabular}{ccccccccccccc}
\toprule
\multirow{3}{*}{Steganography Methods} & \multicolumn{12}{c}{Relative Payloads (bps)}\\
\cmidrule(lr){2-13}
  & 0.1 & 0.2 & 0.3 & 0.4 & 0.5 & 1 & 0.1 & 0.2 & 0.3 & 0.4 & 0.5 & 1\\
\cmidrule(lr){2-7} \cmidrule(lr){8-13}
  & \multicolumn{6}{c}{TIMIT} & \multicolumn{6}{c}{LJSpeech} \\
\midrule
DFR \cite{chen2019derivative}& 41.77 & 39.29 & 31.48 & 27.05 & 23.76 & 14.92 & 37.34 & 32.99 & 28.68 & 21.74 & 16.25 & 10.21 \\
ACC \cite{8}& 42.10 & 40.08 & 35.11 & 31.04 & 26.93 & 18.77 & 39.88 & 36.76 & 31.08 & 28.66 & 22.98 & 19.73 \\
IAA\_Flat \cite{12}& 49.04 & 47.89 & 46.66 & 44.67 & 40.63 & 35.81 & 46.21 & 44.83 & 40.07 & 37.62 & 33.84 & 29.71 \\
GAIE\_MAS \cite{10}& \textbf{49.89} & 49.01 & 48.52 & 47.68 & 46.77 & 44.12 & 49.73 & 48.97 & 48.32 & 47.41 & 46.26 & 44.22 \\
Hifi-Stego \cite{11007018}& 33.21 & 21.98 & 10.23 & - & - & - & 29.22 & 17.78 & 8.84 & - & - & - \\
FGAS (Ours) & 49.73 & \textbf{49.26} & \textbf{48.92} & \textbf{48.40} & \textbf{47.76} & \textbf{45.01} & \textbf{49.86} & \textbf{49.25} & \textbf{48.88} & \textbf{47.79} & \textbf{46.35} & \textbf{45.90} \\
\midrule
& \multicolumn{6}{c}{GTZAN} & \multicolumn{6}{c}{Audioset} \\
\midrule
DFR \cite{chen2019derivative}& 43.15 & 40.52 & 33.64 & 30.91 & 25.10 & 16.55 & 39.82 & 35.14 & 32.22 & 27.45 & 25.30 & 19.15 \\
ACC \cite{8}& 45.50 & 43.18 & 39.65 & 36.40 & 31.85 & 27.10 & 44.15 & 41.60 & 39.25 & 37.05 & 30.40 & 23.88 \\
IAA\_Flat \cite{12}& 49.12 & 48.95 & 47.30 & 45.82 & 42.15 & 38.60 & 48.10 & 46.55 & 42.30 & 39.15 & 35.60 & 31.45 \\
GAIE\_MAS \cite{10}& 49.05 & 49.30 & 48.95 & 48.20 & 47.15 & 45.30 & 49.58 & 49.15 & 48.60 & 47.95 & 46.80 & 44.90 \\
Hifi-Stego \cite{11007018}& 27.59 & 19.88 & 7.62 & - & - & - & 30.52 & 19.17 & 10.28 & - & - & - \\
FGAS (Ours) & \textbf{49.92} & \textbf{49.55} & \textbf{49.21} & \textbf{48.68} & \textbf{47.53} & \textbf{46.85} & \textbf{49.94} & \textbf{49.60} & \textbf{49.13} & \textbf{48.55} & \textbf{47.40} & \textbf{46.52} \\
\bottomrule
\end{tabular}
\end{table*}
\begin{table*}[!t]
\centering
\caption{The Anti-Steganalysis Performance $\overline{P}_{E}$ (IN \%) of The Tested Steganography methods in Resisting The Detection of \textbf{LinNet} Steganalyzer}
\label{tab:LinNetResults}
\begin{tabular}{ccccccccccccc}
\toprule
\multirow{3}{*}{Steganography Methods} & \multicolumn{12}{c}{Relative Payloads (bps)}\\
\cmidrule(lr){2-13}
  & 0.1 & 0.2 & 0.3 & 0.4 & 0.5 & 1 & 0.1 & 0.2 & 0.3 & 0.4 & 0.5 & 1\\
\cmidrule(lr){2-7} \cmidrule(lr){8-13}
  & \multicolumn{6}{c}{TIMIT} & \multicolumn{6}{c}{LJSpeech} \\
\midrule
DFR \cite{chen2019derivative} & 38.61 & 35.23 & 32.48 & 26.55 & 23.68 & 14.33 & 37.37 & 34.01 & 28.70 & 23.65 & 15.88 & 11.10 \\
ACC \cite{8}& 45.75 & 40.09 & 37.33 & 34.22 & 31.36 & 20.88 & 43.62 & 37.88 & 34.01 & 29.82 & 21.67 & 17.03 \\
IAA\_Flat \cite{12}& 44.19 & 41.75 & 39.96 & 39.09 & 37.90 & 29.83 & 43.66 & 41.57 & 37.63 & 35.10 & 31.62 & 27.34 \\
GAIE\_MAS \cite{10} & 46.91 & 46.33 & 45.37 & 44.85 & 44.08 & 41.94 & 45.64 & 45.28 & 44.30 & 43.96 & 42.81 & 41.03 \\
Hifi-Stego \cite{11007018}& 27.67 & 23.84 & 11.25 & - & - & - & 26.37 & 19.22 & 7.09 & - & - & - \\
FGAS (Ours) & \textbf{47.37} & \textbf{46.99} & \textbf{45.62} & \textbf{44.93} & \textbf{44.51} & \textbf{42.26} & \textbf{46.44} & \textbf{45.82} & \textbf{44.93} & \textbf{44.04} & \textbf{43.11} & \textbf{41.89} \\
\midrule
& \multicolumn{6}{c}{GTZAN} & \multicolumn{6}{c}{Audioset} \\
\midrule
DFR \cite{chen2019derivative} & 40.25 & 37.60 & 34.15 & 28.30 & 25.10 & 16.20 & 39.10 & 35.80 & 30.15 & 25.40 & 22.55 & 16.80 \\
ACC \cite{8}& 47.10 & 42.35 & 39.10 & 36.55 & 33.20 & 29.45 & 43.20 & 40.75 & 39.10 & 33.50 & 28.90 & 20.15 \\
IAA\_Flat \cite{12}& 46.30 & 45.85 & 41.50 & 40.25 & 39.10 & 31.60 & 45.50 & 43.20 & 39.40 & 36.85 & 33.15 & 29.20 \\
GAIE\_MAS \cite{10} & 47.85 & 47.20 & 46.55 & 45.90 & 45.15 & 43.20 & 47.10 & 46.50 & 45.85 & 44.90 & 43.75 & 42.10 \\
Hifi-Stego \cite{11007018}& 31.21 & 23.88 & 12.71 & - & - & - & 28.77 & 19.34 & 9.65 & - & - & - \\
FGAS (Ours) & \textbf{48.50} & \textbf{47.95} & \textbf{47.10} & \textbf{46.32} & \textbf{45.85} & \textbf{44.15} & \textbf{47.80} & \textbf{47.25} & \textbf{46.43} & \textbf{45.60} & \textbf{44.55} & \textbf{43.39} \\
\bottomrule
\end{tabular}
\end{table*}
\begin{table*}[t]
\centering
\caption{The Anti-Steganalysis Performance $\overline{P}_{E}$ (IN \%) of The Tested Steganography methods in Resisting The Detection of TMF Steganalyzer}
\label{tab:TMFresults}
\begin{tabular}{ccccccccccccc}
\toprule
  \multirow{3}{*}{Steganography Methods} & \multicolumn{12}{c}{Relative Payloads (bps)}\\
\cmidrule(lr){2-13}
  & 0.1 & 0.2 & 0.3 & 0.4 & 0.5 & 1 & 0.1 & 0.2 & 0.3 & 0.4 & 0.5 & 1\\
\cmidrule(lr){2-7} \cmidrule(lr){8-13}
  & \multicolumn{6}{c}{TIMIT} & \multicolumn{6}{c}{LJSpeech} \\
\midrule
DFR \cite{chen2019derivative} &45.32&40.63&36.47& 32.90&29.27&19.88&45.14&38.71&32.15 &25.92&20.07&15.73\\
ACC \cite{8} & 47.73& 44.81 & 42.60 &39.92 & 37.06 &31.13 & 45.40 & 43.78 &30.60 & 37.76 & 33.58 &25.33 \\
IAA\_Flat \cite{12} &49.41 &48.87& 48.28 & 48.01 & 47.30&39.22  & 49.26 & 47.02 &\textbf{46.38} & 38.65& 36.69 & 29.77\\
GAIE\_MAS \cite{10} & \textbf{49.95} & \textbf{49.74} & \textbf{49.54} &\textbf{49.20} & 48.80 & 41.47 & \textbf{49.31} & 47.69 & 45.06 &  41.50& 37.13 & 33.98 \\
Hifi-Stego\cite{11007018}&31.23&20.67&13.30&-&-& -&28.17&17.88&10.09&-& -&-\\
FGAS (Ours) &  49.53 & 49.46 & 49.30 & 49.19 & \textbf{48.87} & \textbf{42.01} & 49.22 & \textbf{47.71} & 45.84 & \textbf{43.57}& \textbf{39.62} & \textbf{35.55} \\
\midrule
& \multicolumn{6}{c}{GTZAN} & \multicolumn{6}{c}{Audioset} \\
\midrule
DFR \cite{chen2019derivative} & 46.19 & 43.21 & 39.50 & 35.15 & 31.43 & 27.38 & 43.56 & 42.85 & 37.62 & 32.17 & 28.48 & 23.90 \\
ACC \cite{8}  & 47.92 & 46.35 & 44.10 & 41.50 & 38.80 & 34.20 & 46.18 & 43.64 & 39.25 & 35.82 & 30.53 & 27.17 \\
IAA\_Flat \cite{12} & 49.69 & 49.21 & 48.84 & 48.56 & 47.78 & 41.23   & 48.20 & 47.88 & 46.50 & 45.27 & 43.40 & 39.15 \\
  GAIE\_MAS \cite{10} & 49.88 &\textbf{49.53} & 49.02 & \textbf{48.95} & \textbf{48.40} & 42.80 & 49.75 & 49.22 & 48.15 &  \textbf{47.43} & \textbf{45.22} & 41.50 \\
Hifi-Stego\cite{11007018}&28.55&17.03&9.78&-&-&-&33.45&21.57&12.26&-&-&-\\
FGAS (Ours) &  \textbf{49.90} & 49.44 & \textbf{49.15} & 48.80 & 48.25 & \textbf{44.60} &  \textbf{49.85} & \textbf{49.37} & \textbf{48.40} &  47.28 & 44.85 & \textbf{42.47}  \\
\bottomrule
\end{tabular}
\end{table*}
\subsection{Robustness}
\subsubsection{Robustness under non-attack conditions} Table ~\ref{tab:robustnesswithout} reports the secret message extraction accuracy of various steganography methods under non-attack (Clean) conditions. It is worth noting that traditional steganographic schemes, including DFR~\cite{chen2019derivative}, ACC~\cite{8}, IAA\_flat~\cite{12}, and GAIE\_MAS~\cite{10}, employ coding-based embedding frameworks such as STCs or Wet Paper Coding (WPC). According to the design philosophy of these coding techniques~\cite{1597139}, error-free extraction is mathematically guaranteed provided the payload does not exceed the channel capacity. Consequently, their accuracy under the Clean condition is theoretically 100\% and is denoted as "-" in the table for conciseness.\\
\footnotetext[3]{https://www.kaggle.com/datasets/andradaolteanu/gtzan-dataset-music-genre-classification}
\footnotetext[4]{https://research.google.com/audioset/}
\indent In contrast, the proposed FGAS achieves an extraction accuracy close to 100\% without any auxiliary error-correction coding. This demonstrates that the A\textsuperscript{2}PG strategy effectively aligns the stego audio with the fixed decoder's decision boundaries, establishing a highly reliable communication channel across diverse audio contents.\\
\subsubsection{Robustness with attack conditions}
\indent When transmitted over communication channels, stego audio inevitably encounters various unpredictable disturbances. These attacks compromise the accuracy of secret information extraction, thereby undermining the practical reliability of covert communication systems. This section conducts a comprehensive robustness assessment of several existing audio steganography methods using six typical attacks as examples: random Gaussian noise, MP3 and AAC compression, low-pass filtering, band-pass filtering, and stretching.\\
\indent Table~\ref{tab:robustnesswithout} presents the secret message extraction accuracy rates of different steganography methods under six typical attacks. The standard FGAS, like other traditional audio steganography methods, suffers from low extraction accuracy and lacks sufficient robustness against common signal distortions. This vulnerability stems from the fact that the embedded secret messages rely on extremely minute and precise temporal perturbations for encoding. These perturbations are highly susceptible to degradation under attacks such as noise, compression, and timing alterations, which prevent the fixed decoder from extracting valid features.\\
\indent To address this, the improved FGAS method incorporating the A\textsuperscript{2}PG-R extension successfully identifies robust perturbations that remain stable under these common attacks, thereby ensuring reliable extraction of secret information. To evaluate its performance, we compare the robustness-enhanced FGAS with Hifi-stego \cite{11007018}, a recent DNN-based audio steganography method designed with inherent robustness. As demonstrated in Table~\ref{tab:robustnesswithout}, the FGAS method with the A\textsuperscript{2}PG-R extension achieves superior recovery accuracy across all six common attacks compared to Hifi-stego. Even under aggressive MP3 compression at 32~kbps or lower, FGAS remains competitive in terms of message recovery accuracy. These results highlight that the A\textsuperscript{2}PG-R strategy significantly strengthens the resilience of the generated perturbations against diverse channel distortions.
\subsection{Security}
\subsubsection{Resistance to Audio Steganalyzers}
\indent To comprehensively evaluate the anti-st eganalysis performance of FGAS, we conduct experiments under different relative payloads and assess the performance using three publicly available audio steganalyzers, deep learning steganalyzers ChenNet \cite{23}, LinNet \cite{24}, and traditional feature-based detectors TMF \cite{luo2018improved}. This allows us to effectively analyze and compare the anti-steganalysis performance of each method.\\
\indent It should be noted that Hifi-Stego~\cite{11007018} adopts "bits per second" as its payload metric, which we have normalized to "bits per sample" (bps) for a fair comparison. Specifically, by adjusting its hyper-parameter $\sigma \in \{1, 2, 3\}$, Hifi-Stego achieves maximum capacities of 2048, 4096, and 6144 bits per second, respectively. Given the 16 kHz sampling rate of the datasets, these correspond to relative payloads of approximately 0.128, 0.256, and 0.384 bps. Consequently, we report its performance at 0.1, 0.2, and 0.3 bps using the corresponding $\sigma$ settings. For relative payloads exceeding 0.4 bps, Hifi-Stego is denoted as "-" due to its intrinsic capacity constraints.\\
\begin{table}[t]
    \centering
    \caption{AUROC (\%) of four statistical detection methods
    for distinguishing FGAS stego audio from cover audio
    on four datasets.}
    \label{tab:statistical_adversarial_detection}
    \small
    \renewcommand{\arraystretch}{1.15}
    \setlength{\tabcolsep}{2pt}
    \begin{tabular*}{\columnwidth}
        {@{\extracolsep{\fill}}lcccc@{}}
        \toprule
        Method & TIMIT & LJSpeech & GTZAN & AudioSet \\
        \midrule
    PCA~\cite{hendrycks2016early}& 49.96 & 50.11 & 50.32 & 50.24 \\
    KDE~\cite{feinman2017detecting} & 49.79 & 49.83 & 50.15 & 50.38 \\
LID~\cite{ma2018characterizing}& 50.21 & 49.67 & 49.80 & 50.52 \\
        Mahalanobis distance~\cite{lee2018simple} & 49.89 & 50.08 & 49.94 & 50.06 \\
        \bottomrule
    \end{tabular*}
\end{table}
\begin{table}[t]
\centering
\caption{The Average Stego Audio Quality Metric of Different Steganography Methods Tested on Speech and Music Datasets}
\label{tab:2}
\setlength{\tabcolsep}{4pt}
\begin{tabular}{@{}cccccc@{}}
\toprule
Steganography & \multirow{2}{*}{metric} & \multirow{2}{*}{TIMIT} & \multirow{2}{*}{LJSpeech} & \multirow{2}{*}{GTZAN} & \multirow{2}{*}{Audioset} \\
Methods & & & & & \\
\midrule
\multirow{2}{*}{DFR \cite{chen2019derivative}}
& PSNR$\uparrow$ & 103.45 & 100.52 & 98.10 & 101.30 \\
& PEAQ$\uparrow$ & 4.2261 & 4.2146 & 4.1713 & 4.2002 \\
\midrule
\multirow{2}{*}{ACC \cite{8}}
& PSNR$\uparrow$ & 105.33 & 104.27 & 106.34 & 105.11 \\
& PEAQ$\uparrow$ & 4.4655 & 4.3825 & 4.4209 & 4.4637 \\
\midrule
\multirow{2}{*}{IAA\_flat \cite{12}}
& PSNR$\uparrow$ & 106.72 & 107.20 & 108.54 & 107.29 \\
& PEAQ$\uparrow$ & 4.3728 & 4.3544 & 4.4251 & 4.4017 \\
\midrule
\multirow{2}{*}{GAIE\_MAS \cite{10}}
& PSNR$\uparrow$ & 98.39 & 97.27 & 99.14 & 99.82 \\
& PEAQ$\uparrow$ & 4.5370 & 4.5301 & 4.5404 & 4.5413 \\
\midrule
\multirow{2}{*}{Hifi-Stego \cite{11007018}}
& PSNR$\uparrow$ & 26.54 & 19.85 & 20.18 & 27.62 \\
& PEAQ$\uparrow$ & 2.6711 & 1.9606 & 2.3321 & 2.6925 \\
\midrule
\multirow{2}{*}{FGAS (Ours)}
& PSNR$\uparrow$ & \textbf{107.94} & \textbf{107.56} & \textbf{109.63} & \textbf{108.17} \\
& PEAQ$\uparrow$ & \textbf{4.5415} & \textbf{4.5403} & \textbf{4.5432} & \textbf{4.5420} \\
\bottomrule
\end{tabular}
\end{table}
\indent The results of resisting the detection of ChenNet and LinNet are shown in Table~\ref{tab:ChenNetResults} and Table~\ref{tab:LinNetResults}. Across all relative payload ranges from 0.1 bps to 1.0 bps, our method maximizes the classification error rate ($\overline{P}_{E}$), bringing the steganalysis accuracy of ChenNet and LinNet close to random guessing. Compared to other methods, our method has higher values of $\overline{P}_{E}$ on both speech and music datasets, with less decline at high relative payloads. This demonstrates that FGAS is well adapted to different audio content and statistical properties. The results shown in Table~\ref{tab:TMFresults} demonstrate that despite TMF not participating in the perturbation optimization process, FGAS maintains comparable anti-steganalysis performance to SOTA methods. More compellingly, as the load increases, FGAS achieves the best $\overline{P}_{E}$ values across all four datasets in both the 0.5 bps and 1 bps.\\
\indent The superior anti-steganalysis performance results from incorporating anti-steganalysis feedback into perturbation optimization. This high-level design introduces a specialized anti-steganalysis loss, strategically incorporated during the later stages of perturbation optimization. The $\gamma$ parameter facilitates the precise trade-off between steganographic quality and detection resistance. By iteratively optimizing the perturbation using steganalysis feedback, the resulting stego audio is statistically optimized to mimic the cover audio distribution, thereby significantly hindering the steganalysers' classification accuracy.
\subsubsection{Resistance to Statistical Adversarial Example Detectors}
To evaluate the concealment of FGAS against statistical
adversarial example detection methods, we conduct cover-versus-stego
detection experiments on TIMIT, LJSpeech, GTZAN, and AudioSet.
We consider four methods discussed in the review by Zhao et
al.~\cite{10380989}: Principal Component Analysis (PCA)~\cite{hendrycks2016early},
Kernel Density Estimation (KDE)~\cite{feinman2017detecting}, Local Intrinsic Dimensionality
(LID)~\cite{ma2018characterizing}, and Mahalanobis distance~\cite{lee2018simple}. Each method is evaluated on
its ability to distinguish FGAS stego audio from cover audio
using the area under the receiver operating characteristic
curve (AUROC), where 50\% represents random-level discrimination, as shown in Table~\ref{tab:statistical_adversarial_detection}.\\
\indent The AUROC values of all four methods remain close to 50\%
across the four datasets, ranging from 49.67\% to 50.52\%.
The maximum absolute deviation from 50\% is only 0.52
percentage points. These results indicate that the evaluated
statistical detection methods exhibit limited ability to
distinguish FGAS stego audio from cover audio under the
tested settings.
\begin{figure}[t]
\centering
\includegraphics[width=8.8cm]{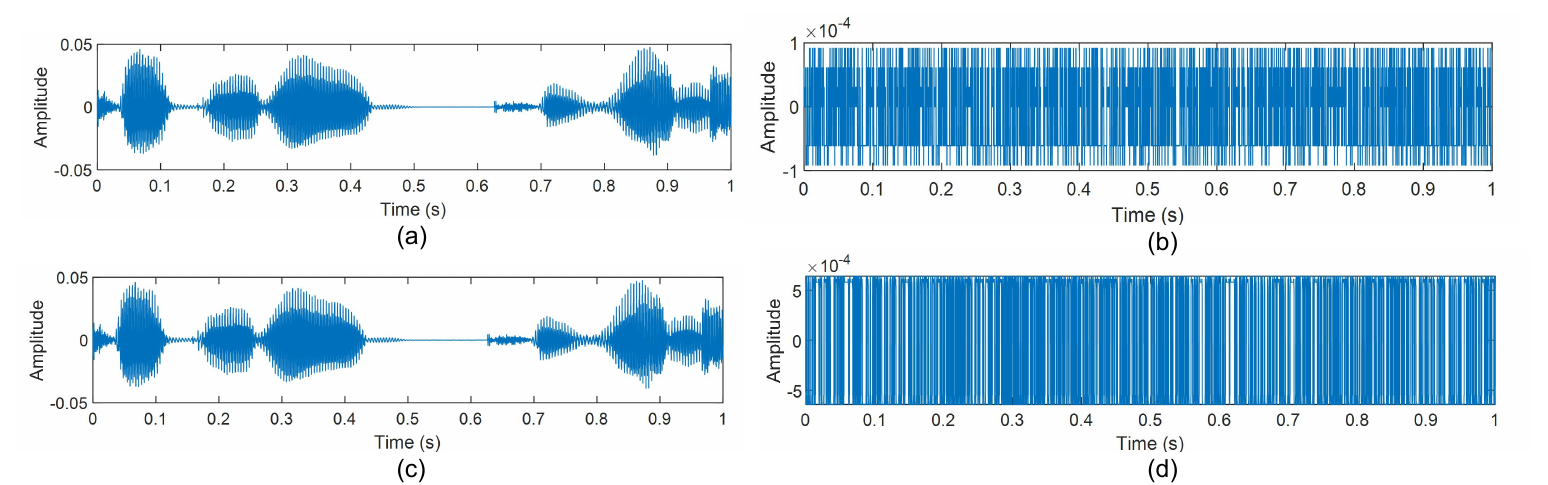}
\caption{Time-domain waveforms of TIMIT audio clips and perturbations, illustrating the effect of the anti-steganalysis module on perturbation amplitude over time. (a): Cover audio waveform. (b): Waveforms of the perturbation generated with the anti-steganalysis module. (c): Stego audio waveform. (d): Waveforms of the perturbation generated without the anti-steganalysis module.}
\label{fig:time}
\end{figure}
\begin{figure}[!t]
\centering
\includegraphics[width=8.8cm]{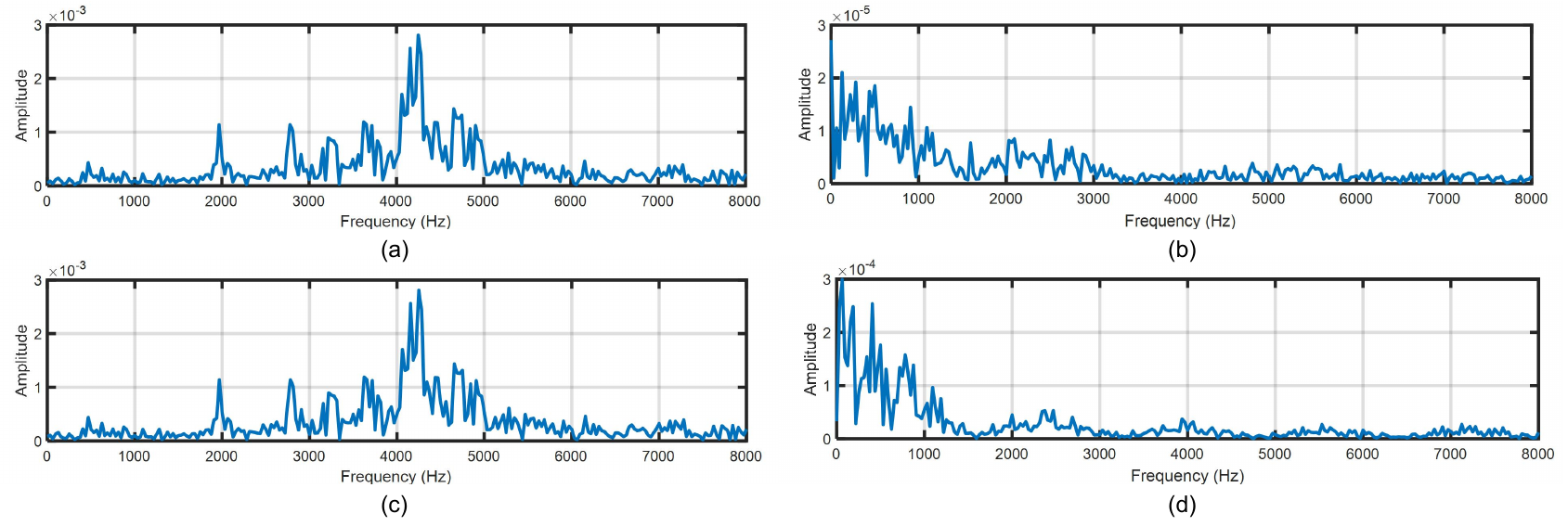}
\caption{Frequency spectra of TIMIT audio clips and perturbations, illustrating the effect of the anti-steganalysis module on the spectral distribution of perturbations. (a): Cover audio waveform. (b): Waveforms of the perturbation generated with the anti-steganalysis module. (c): Stego audio waveform. (d): Waveforms of the perturbation generated without the anti-steganalysis module.}
\label{fig:wav}
\end{figure}
\begin{figure}[!t]
\centering
\includegraphics[width=8.8cm]{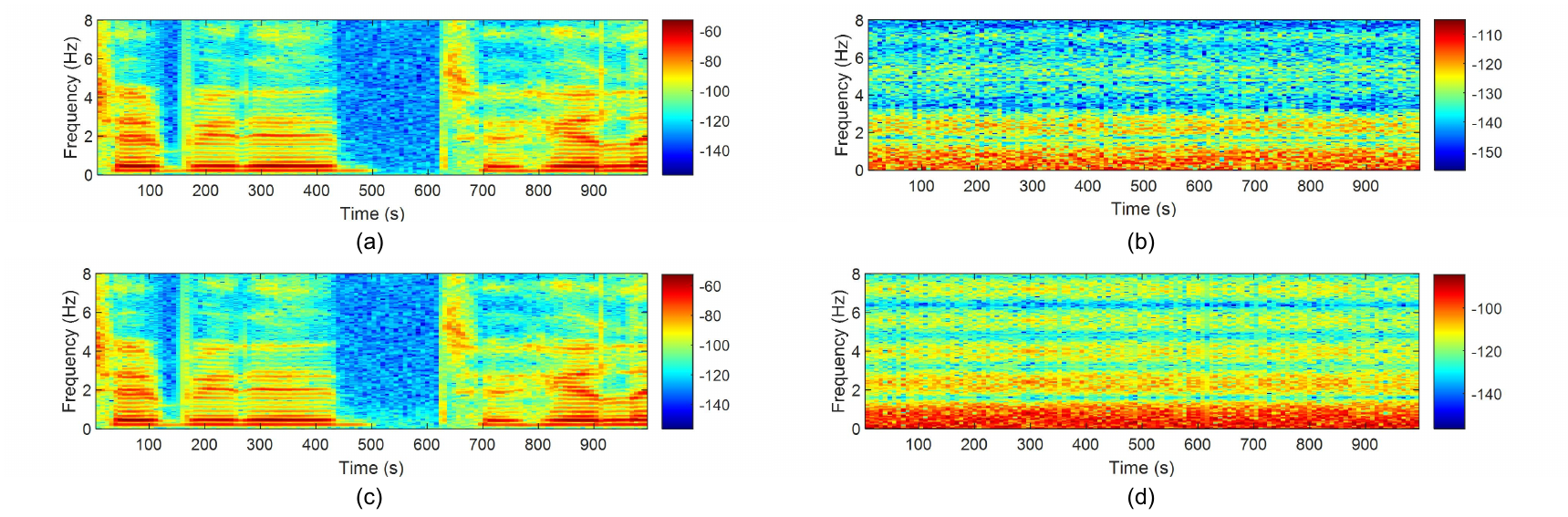}
\caption{Energy maps of TIMIT audio clips and perturbations, illustrating the effect of the anti-steganalysis module on the distribution of perturbation energy. (a): Cover audio energy map. (b): Energy map of the perturbation generated with the anti-steganalysis module. (c): Stego audio energy map. (d): Energy map of the perturbation generated without the anti-steganalysis module.}
\label{fig:power}
\end{figure}
\subsection{Stego Audio Quality}
\indent The stego audio quality is a critical metric, as it directly determines the imperceptibility of hidden data and significantly influences the resistance against modern steganalysis. Table~\ref{tab:2} provides a quantitative comparison of different methods using PSNR and PEAQ. Our proposed FGAS consistently achieves superior performance across diverse speech and music datasets. This exceptional transparency is attributed to two key design strategies. First, the A\textsuperscript{2}PG strategy incorporates a strictly constrained optimization objective, where the steganographic loss functions work in tandem with the $\epsilon$-bound to limit the maximum amplitude of adversarial perturbations. Unlike traditional additive noise, these perturbations are adaptively concentrated in regions that are less sensitive to the Human Auditory System (HAS). Second, the application of the $\tanh$ function acts as a differentiable soft-thresholding operator. This ensures that the generated perturbations are not only bounded in magnitude but also smoothly distributed in the temporal domain, avoiding abrupt signal changes that could be detected by high-order statistical steganalysis. Consequently, FGAS maintains high structural similarity to the cover audio, rendering the stego signals virtually indistinguishable from their original counterparts in both subjective listening and objective statistical tests.
\begin{table*}[t]
\centering
\caption{The Anti-Steganalysis Performance $\overline{P}_{E}$ (IN \%) With and Without Anti-steganalysis Module Tested on Speech and Music Datasets in Resisting The Detection of ChenNet and LinNet Steganalyzer}
\label{tab:4}
\begin{tabular}{cccccccccccccc}
\toprule
\multirow{3}{*}{Steganalyzer} & \multirow{3}{*}{Module} & \multicolumn{12}{c}{Relative Payloads (bps)}\\
\cmidrule(lr){3-14}
&  & 0.1 & 0.2 & 0.3 & 0.4 & 0.5 & 1 &0.1 & 0.2 & 0.3 & 0.4 & 0.5 & 1 \\
\cmidrule(lr){3-8} \cmidrule(lr){9-14}
&  & \multicolumn{6}{c}{TIMIT} &  \multicolumn{6}{c}{LJSpeech} \\
\midrule
\multirow{2}{*}{ChenNet~\cite{23}}
& Without &39.81& 29.73 & 26.76 & 25.52 & 24.53& 15.34 & 40.17 &31.04& 25.50 &23.11&20.08& 15.56\\
    & With&\textbf{49.73} & \textbf{49.26} & \textbf{48.92} & \textbf{48.40} & \textbf{47.76} & \textbf{45.01}& \textbf{49.86} & \textbf{49.05} & \textbf{48.88} & \textbf{47.79} & \textbf{46.35} & \textbf{45.90} \\
\midrule
\multirow{2}{*}{LinNet~\cite{24}}
&Without& 38.14 & 25.03 &22.62&21.71&23.55&13.87
&36.20 &23.96 & 20.07& 28.84 & 17.61 & 13.99\\
& With& \textbf{47.37} & \textbf{46.99} & \textbf{45.62} & \textbf{44.93} & \textbf{44.21} & \textbf{42.07}& \textbf{46.44} & \textbf{45.82} & \textbf{44.93} & \textbf{44.04} & \textbf{43.11} & \textbf{41.89} \\
\midrule
&& \multicolumn{6}{c}{GTZAN} & \multicolumn{6}{c}{Audioset} \\
\midrule
\multirow{2}{*}{ChenNet~\cite{23}}
& Without &37.62& 32.30 & 27.99 & 26.68 & 23.21& 17.45 & 38.25 &33.11& 28.30 &24.78&21.94& 16.33\\
    & With&\textbf{49.92} & \textbf{49.55} & \textbf{49.21} & \textbf{48.68} & \textbf{47.53} & \textbf{46.85}& \textbf{49.94} & \textbf{49.60} & \textbf{49.13} & \textbf{48.55} & \textbf{47.41} & \textbf{46.52} \\
\midrule
\multirow{2}{*}{LinNet~\cite{24}}
&Without& 38.26 & 34.90&25.44&21.69&19.71&14.58
&39.04 &33.13 & 24.85& 23.84 & 15.17 & 11.45\\
& With& \textbf{48.50} & \textbf{47.95} & \textbf{47.10} & \textbf{46.32} & \textbf{45.85} & \textbf{44.15}& \textbf{47.84} & \textbf{47.25} & \textbf{46.43} & \textbf{45.60} & \textbf{44.55} & \textbf{43.39} \\
\bottomrule
\end{tabular}
\label{tab:absteg}
\end{table*}
\begin{table}[t]
\centering
\caption{The Stego Audio Quality  With and Without Anti-steganalysis Module Tested on Speech and Music Datasets in Resisting The Detection of ChenNet and LinNet Steganalyzer}
\begin{tabular}{cccccc}
\toprule
Module &metric&TIMIT & LJSpeech &GTZAN&Audioset \\ 
\midrule
\multirow{2}{*}{Without}&PSNR$\uparrow$ &107.86&107.37&109.48&107.85 \\
&PEAQ$\uparrow$ & 4.54&4.54&4.54&4.54 \\
\midrule
\multirow{2}{*}{With}&PSNR$\uparrow$ &\textbf{107.94}&\textbf{107.56}&\textbf{109.63}&\textbf{108.17} \\
&PEAQ$\uparrow$ & 4.54&4.54&4.54&4.54  \\
\bottomrule
\end{tabular}
\label{tab:abquality}
\end{table}

\subsection{Generalization ability}
\indent To further evaluate the generalization capability of the proposed FGAS scheme under unseen steganalyzer conditions, we introduce a variant called FGAS*. Unlike the original FGAS setup, which leverages a specific steganalyzer for both adversarial perturbation generation and evaluation, FGAS* employs a mismatched steganalyzer setup: perturbations are generated with one steganalyzer and evaluated with another. This simulates real-world scenarios in which the steganalyzer architecture is unknown at embedding time, enabling us to assess the transferability of the generated perturbations across detection models.\\
\indent The results, presented in Table~\ref{tab:generalization ability speech}, demonstrate that FGAS* maintains high $\overline{P}_{E}$ values during the evaluation of the cross-model, closely matching the performance of the standard FGAS. This suggests that the learned perturbations are highly transferable and are not overfit to a specific steganalysis model, thus confirming the practical applicability of FGAS under unknown detection environments.\\
\begin{table*}[t]
\centering
\caption{The Anti-Steganalysis Performance $\overline{P}_{E}$ (IN \%) of FGAS and FGAS* Tested on Speech and Music Datasets in Resisting The Detection of ChenNet and LinNet Steganalyzer}
\label{tab:generalization ability speech}
\begin{tabular}{cccccccccccccc}
\toprule
\multirow{3}{*}{Steganalyzer} & \multirow{3}{*}{Steganography Methods} & \multicolumn{12}{c}{Relative Payloads (bps)}\\
\cmidrule(lr){3-14}
&  & 0.1 & 0.2 & 0.3 & 0.4 & 0.5 & 1 & 0.1 & 0.2 & 0.3 & 0.4 & 0.5 & 1\\
\cmidrule(lr){3-8} \cmidrule(lr){9-14}
&  & \multicolumn{6}{c}{TIMIT} & \multicolumn{6}{c}{LJSpeech} \\
\midrule
\multirow{2}{*}{ChenNet}& FGAS & \textbf{49.73} & \textbf{49.26} &\textbf{48.92} & \textbf{48.40} & \textbf{47.76} & \textbf{45.01}& \textbf{49.86} & \textbf{49.25} & \textbf{48.88} & \textbf{47.79} & 46.35 & \textbf{45.90} \\
&FGAS\textsuperscript{*} & 49.55 & 48.73 & 48.36 & 47.89 & 47.02 & 44.73 & 49.65 & 48.89 & 48.61 & 47.32 & \textbf{46.54} & 45.09 \\
\midrule
\multirow{2}{*}{LinNet}& FGAS& \textbf{47.37} & \textbf{46.99} & \textbf{45.62} & \textbf{44.93} & \textbf{44.51} & \textbf{42.26}& \textbf{46.44} & \textbf{45.82}& \textbf{44.93} & \textbf{44.04}& \textbf{43.11} &\textbf{41.89} \\
&FGAS\textsuperscript{*} & 47.03 & 46.52 & 45.41 & 44.67 & 44.12 & 42.07& 46.02 & 45.58 & 44.52 & 43.61 & 43.04 & 41.62\\
\midrule
&& \multicolumn{6}{c}{GTZAN} & \multicolumn{6}{c}{Audioset} \\
\midrule
\multirow{2}{*}{ChenNet}
& FGAS & \textbf{49.92} & \textbf{49.55} & \textbf{49.21}& \textbf{48.68} & \textbf{47.53} &\textbf{46.85} & \textbf{49.94} & \textbf{49.60} & \textbf{49.15} & \textbf{48.55} & \textbf{47.40} & \textbf{46.52} \\
&FGAS\textsuperscript{*} &49.06&48.71&48.30&47.86&47.04&45.88&48.22&48.01&47.56&46.99&46.13&45.84\\
\midrule
\multirow{2}{*}{LinNet}
& FGAS& \textbf{48.50} & \textbf{47.95} & \textbf{47.10} & \textbf{46.37} & \textbf{45.85} & 44.15 &\textbf{47.80} & \textbf{47.25} & \textbf{46.40} & \textbf{45.60} & \textbf{44.55} & \textbf{43.30} \\
&FGAS\textsuperscript{*} &48.34&47.23&46.66&45.90&45.37&\textbf{44.28}&47.02&46.77&46.21&45.43&44.27&43.06\\

\bottomrule
\end{tabular}
\end{table*}
\begin{table*}[t]
\caption{Comparison of secret message recovery accuracy (\%) between static attack mixture and curriculum learning under different attacks at a relative payload of 0.1 bps.}
\label{tab:curriculum_ablation}
\centering
\small
\setlength{\tabcolsep}{3pt}
\begin{tabular}{@{}l*{14}{c}@{}}
\toprule
\multirow{2}{*}{Method}
& \multicolumn{3}{c}{Noise}
& \multicolumn{5}{c}{MP3 compression (kbps)}
& AAC & LP-F & BP-F
& \multicolumn{3}{c}{Stretch} \\
\cmidrule(lr){2-4}   \cmidrule(lr){5-9}
\cmidrule(lr){10-10} \cmidrule(lr){11-11}
\cmidrule(lr){12-12} \cmidrule(lr){13-15}
& 20 dB & 30 dB & 50 dB
& 16 & 32 & 64 & 128 & 256
& 256 kbps & 4k & 0.3--8k & 0.8 & 0.9 & 1.1 \\
\midrule
Static Attack Mixture
& 77.97 & 88.52 & 94.09
& 65.77 & 72.21 & 82.13 & 86.31 & 89.50
& 86.41 & 62.24 & 68.71
& 88.13 & 92.96 & 92.14 \\
\midrule
Curriculum Learning
& \textbf{79.12} & \textbf{91.31} & \textbf{94.44}
& \textbf{66.26} & \textbf{78.03} & \textbf{88.75}
& \textbf{89.69} & \textbf{89.75}
& \textbf{87.44} & \textbf{65.25} & \textbf{74.12}
& \textbf{94.18} & \textbf{94.99} & \textbf{93.44} \\
\bottomrule
\end{tabular}
\label{tab:curriculum learning}
\end{table*}
\subsection{Ablation Studies}
\begin{table}[t]
\centering
\caption{Computational Complexity and Generation Efficiency of Various Audio Steganography Methods.}
\begin{tabular}{ccc}
\toprule
Method & Pre-training & Generation Time(s)\\
\midrule
DFR \cite{chen2019derivative} & $\backslash$  & 0.3104\\
ACC \cite{8} &  $\backslash$  & 0.1389\\
IAA\_flat \cite{12} &  $\backslash$ & 0.0378\\
GAIE-MAS \cite{10} & $\backslash$ & 0.0162\\
Hifi-Stego \cite{11007018}&3000 epochs&0.2162\\
FGAS (Ours)&  $\backslash$  & 8.2086\\
FGAS\_R (Ours) &  $\backslash$  & 35.1032\\
\bottomrule
\end{tabular}
\label{tab:timeperformance_transposed_2}
\end{table}
\begin{table}[t]
\centering
\caption{Comparison of Communication Overhead and Deployment Requirements for Different Audio Steganography Methods.}
\label{tab:timeperformance_transposed_1_makecell}
\begin{tabular}{cccc}
\toprule
Method & \makecell{Complex \\ Algorithm}
& Secret Key & \makecell{Complex \\ network \\ architecture} \\
\midrule
DFR \cite{chen2019derivative} & $\surd$ & Protocol Parameters & - \\
ACC \cite{8} & $\surd$ & Generator Seed & - \\
IAA\_flat \cite{12} & $\surd$ & Hyperparameters & $\surd$ \\
GAIE-MAS \cite{10} & $\surd$ & Generator Seed & - \\
Hifi-Stego \cite{11007018} & $\surd$ & Mapping Parameters & $\surd$ \\
FGAS (Ours) & $\times$ & Initialization Seed & $\times$ \\
\bottomrule
\end{tabular}

\vspace{3pt}
\begin{minipage}{\columnwidth}
\footnotesize
\textit{Note:} $\surd$ denotes required, $\times$ denotes not required,
and $-$ indicates not applicable.
\end{minipage}
\end{table}
\indent In order to investigate the extent to which the anti-steganalysis module in FGAS contributes to the anti-steganalysis performance and quality of the generated stego audio, we conducted ablation experiments. 
\subsubsection{Effect of the Anti-Steganalysis Module}
The anti-steganalysis module uses gradient feedback from pre-trained
steganalyzers to refine the perturbation at each iteration of
A\textsuperscript{2}PG. Table~\ref{tab:absteg} compares FGAS with and
without this feedback on the speech and music datasets at different
relative payloads. With the module enabled, both ChenNet and LinNet
exhibit higher detection error rates ($\overline{P}_{E}$), indicating
that they have greater difficulty distinguishing stego audio from
cover audio.

The module also improves stego audio quality. As shown in
Table~\ref{tab:abquality}, enabling gradient feedback yields higher PSNR
and PEAQ values. The time-domain waveforms in Fig.~\ref{fig:time},
the spectra in Fig.~\ref{fig:wav}, and the energy maps in
Fig.~\ref{fig:power} illustrate this difference. In the examples shown,
the perturbations generated with gradient feedback have smaller
magnitudes and are less visible against the cover signal.
\subsubsection{Effect of Decoder Kernel Sizes}

Table~\ref{tab:kernel_ablation} compares three decoder configurations on TIMIT. Mixed uses kernels of size~5 in the first two convolutional
layers and size~3 in the last three. All~3 and All~5 use
kernels of size~3 and size~5, respectively, in all five layers.

Mixed achieves the highest PSNR and PESQ at all six tested
payloads. At 0.1 and 0.2~bps, it matches the 99.99\% message recovery
accuracy of All~3. At every tested payload from 0.3 to 1.0~bps,
it achieves higher recovery accuracy than both single-kernel configurations.

The larger kernels in the first two layers of Mixed aggregate
information over a wider local time window. The smaller kernels in the
last three layers then process fine-scale variations in the resulting
features. Using size~3 throughout reduces the decoder's overall
receptive field, whereas using size~5 throughout extends the window
over which each later layer combines neighboring features. The recovery
results suggest that combining wider temporal context in the early layers
with fine-scale processing in the later layers is well suited to decoding
messages embedded in FGAS perturbations.
\subsubsection{Effect of Curriculum Learning}
We compare Static Attack Mixture and Curriculum Learning using the same number of perturbation optimization updates. Static Attack Mixture samples attacks at
random from the full attack set throughout perturbation optimization, whereas Curriculum
Learning introduces simpler attacks first and progressively incorporates
more complex distortions.\\
\indent With the Fixed Decoder Network held constant, Curriculum Learning first
adapts the embedding perturbations to preserve message recovery under
simpler attacks. The resulting perturbations then serve as the starting
point for optimization under more complex distortions. This progression
gives perturbation optimization a more stable starting point at each
successive stage. Under Static Attack Mixture, complex distortions can
affect the updates from the outset.\\
\indent At 0.1~bps, Curriculum Learning achieves higher message recovery accuracy
than Static Attack Mixture under all 14 tested attack conditions
(Table~\ref{tab:curriculum_ablation}). The largest gain occurs under MP3
compression at 64~kbps, where accuracy increases from 82.13\% to 88.75\%,
a gain of 6.62 percentage points. Gains also reach 6.05 percentage points
at a time-stretch factor of 0.8 and 5.41 percentage points under
band-pass filtering.
\subsection{Deployment Efficiency and Embedding Cost}

\subsubsection{Pre-training.}
FGAS and FGAS\_R optimize the perturbation directly on each cover
signal using a fixed decoder. This allows the sender to begin
embedding without first training a model. For a deployment starting
from scratch, this removes the preparation stage that HiFi-Stego
performs over 3,000 epochs
(Table~\ref{tab:timeperformance_transposed_2}).

\subsubsection{One-time deployment.}
The communicating parties share an initialization seed and a
lightweight fixed decoder
(Table~\ref{tab:timeperformance_transposed_1_makecell}).
The same decoder recovers messages from different stego-audio clips,
while all cover-specific optimization takes place at the sender.
Even when the decoder is exchanged only once, its small size limits
both the initial transfer and the model storage needed throughout
operation at the receiver.
\subsubsection{Per-clip embedding.}
FGAS optimizes the embedding perturbation for each cover signal at the
sender, taking 8.2086\,s for a one-second clip. FGAS\_R retains the same
fixed decoder and uses A\textsuperscript{2}PG-R to improve robustness
against signal-processing attacks, taking 35.1032\,s per clip.
Both variants can begin embedding immediately when deployed from scratch.
HiFi-Stego~\cite{11007018}, for comparison, uses a 3{,}000-epoch
training schedule and subsequently embeds a one-second clip in
0.2162\,s. Based on these embedding times, each hour of HiFi-Stego
training avoided offsets the additional embedding time for approximately
450 one-second clips with FGAS or 103 with FGAS\_R, assuming sequential
processing and equal one-time setup costs.
\begin{table}[t]
\centering
\caption{Ablation Study of Mixed-Kernel and Single-Kernel Decoder Configurations on the TIMIT Dataset}
\label{tab:kernel_ablation}
\renewcommand{\arraystretch}{1.15}
\setlength{\tabcolsep}{3pt}
\begin{tabular*}{\columnwidth}{@{\extracolsep{\fill}}ccccc@{}}
\toprule
Payload (bps) & Kernel & PSNR $\uparrow$ & PESQ $\uparrow$ & ACC $\uparrow$ \\
\midrule
\multirow{3}{*}{0.1}
 & Mixed & \textbf{108.12} & \textbf{4.5418} & \textbf{99.99} \\
 & All 3 & 105.26 & 4.5404 & 99.99 \\
 & All 5 & 106.36 & 4.5412 & 98.81 \\
\midrule
\multirow{3}{*}{0.2}
 & Mixed& \textbf{107.57} & \textbf{4.5418} & \textbf{99.99} \\
 & All 3 & 105.41 & 4.5407 & 99.99 \\
 & All 5 & 105.13 & 4.5409 & 95.50 \\
\midrule
\multirow{3}{*}{0.3}
 & Mixed & \textbf{108.07} & \textbf{4.5424} & \textbf{99.90} \\
 & All 3 & 105.93 & 4.5416 & 92.35 \\
 & All 5 & 104.91 & 4.5416 & 95.17 \\
\midrule
\multirow{3}{*}{0.4}
 & Mixed & \textbf{108.02} & \textbf{4.5434} & \textbf{99.50} \\
 & All 3 & 105.41 & 4.5426 & 93.69 \\
 & All 5 & 105.88 & 4.5399 & 94.42 \\
\midrule
\multirow{3}{*}{0.5}
 & Mixed & \textbf{107.73} & \textbf{4.5436} & \textbf{99.95} \\
 & All 3 & 104.74 & 4.5431 & 90.28 \\
 & All 5 & 105.79 & 4.5429 & 95.19 \\
\midrule
\multirow{3}{*}{1.0}
 & Mixed & \textbf{106.83} & \textbf{4.5432} & \textbf{99.10} \\
 & All 3 & 106.07 & 4.5427 & 86.66 \\
 & All 5 & 106.15 & 4.5431 & 93.61 \\
\bottomrule
\end{tabular*}
\label{tab:kernel}
\end{table}

\subsection{Key-Dependent Message Recovery}
Table~\ref{tab:decoder_key} shows how the decoder key affects message
recovery. For each stego signal, we vary the initialization key while
keeping the decoder architecture, input perturbation, and reference
message fixed. At 1 bps, the correct key recovers 99.13\%
of the embedded bits, whereas the incorrect key yields 49.89\%
accuracy. This separation persists across all six payloads: accuracy
with the correct key remains above 99\%, while accuracy with the
tested incorrect keys stays within 1.59 percentage points of random
guessing (50\%).

The decoder accepts \(2^{32}\) distinct initialization seeds. For a
secret seed drawn uniformly from this space, an attacker testing
\(q\) distinct candidates has probability \(q/2^{32}\) of trying the
correct seed. Even after testing one million distinct seeds, this
probability is approximately 0.023\%. Exhaustive enumeration reaches
the correct seed after approximately \(2.15\times10^9\) candidate
trials on average.
\begin{table}[t]
    \centering
    \caption{Bit recovery accuracy (\%) with correct and incorrect
    decoder keys at six relative payloads.}
    \label{tab:decoder_key}
    \small
    \renewcommand{\arraystretch}{1.15}
    \setlength{\tabcolsep}{3.5pt}
    \begin{tabular*}{\columnwidth}{@{\extracolsep{\fill}}lcccccc@{}}
        \toprule
        & \multicolumn{6}{c}{Relative payload (bps)} \\
        \cmidrule(lr){2-7}
        Decoder key & 0.1 & 0.2 & 0.3 & 0.4 & 0.5 & 1.0 \\
        \midrule
        Correct   & 99.99 & 99.99 & 99.95 & 99.55 & 99.78 & 99.13 \\
        Incorrect & 51.38 & 48.41 & 50.21 & 50.38 & 50.65 & 49.89 \\
        \bottomrule
    \end{tabular*}
\end{table}

\section{Conclusion}

This paper presents FGAS, an audio steganography framework that combines a lightweight, key-initialized fixed decoder with per-audio perturbation optimization. The decoder can be independently reconstructed by the communicating parties without encoder or decoder pre-training or the transfer of trained decoder parameters. The proposed Audio Adversarial Perturbation Generation (A\textsuperscript{2}PG) strategy jointly optimizes bounded perturbations for audio fidelity, message recovery, and resistance to steganalysis. Experiments on speech and music datasets demonstrate an average PSNR gain of over 10 dB compared to competing methods, together with accurate message recovery and competitive anti-steganalysis performance against neural and traditional detectors.

Furthermore, A\textsuperscript{2}PG-R incorporates a channel-attack curriculum into perturbation optimization to improve recovery under signal-processing distortions. Together, these results demonstrate the effectiveness of audio-specific fixed decoding and perturbation optimization for high-fidelity audio steganography.
\clearpage
\bibliographystyle{IEEEtran}
\bibliography{references.bib} 
%



\end{document}